\documentclass[final]{aipproc}

\layoutstyle{6x9}

\usepackage{graphicx}

\newcommand{\mhe}{\mathrm{He}}

\newcommand{\hii}{\ion{H}{ii}}
\def\lesssim{\,\lower2truept\hbox{${<\atop\hbox{\raise4truept\hbox{$\sim$}}}$}\,}
\def\gtrsim{\,\lower2truept\hbox{${>\atop\hbox{\raise4truept\hbox{$\sim$}}}$}\,}

\DeclareRobustCommand{\ion}[2]{%
\relax\ifmmode
\ifx\testbx\f@series
{\mathbf{#1\,\mathsc{#2}}}\else
{\mathrm{#1\,\mathsc{#2}}}\fi
\else\textup{#1\,{\mdseries\textsc{#2}}}%
\fi}

\def\druck@lement#1{{}^{\2}_{\3}\mathrm{#1}{}^{\1}_{\4}{}\if@tempswa$\fi} 

\begin{document}

\title{Panchromatic models of galaxies: GRASIL}

\author{Pasquale Panuzzo}{
address={INAF Padova, Vicolo dell'Osservatorio 5, I-35122 Padova, Italy}
}

\author{Laura Silva}{
address={INAF Trieste, Via Tiepolo 11, I-34131 Trieste, Italy}
}

\author{Gian Luigi Granato}{
address={INAF Padova, Vicolo dell'Osservatorio 5, I-35122 Padova, Italy},
altaddress={SISSA, Via Beirut 4, I-34014 Trieste, Italy}
}

\author{Alessandro Bressan}{
address={INAF Padova, Vicolo dell'Osservatorio 5, I-35122 Padova, Italy}
}

\author{Olga Vega}{
address={INAOE, L. E. Erro 1, Tonantzintla, Puebla, Mexico}
}

\begin{abstract}
We present here a model for simulating the panchromatic spectral
energy distribution of galaxies, which aims to be a complete tool
to study the complex multi-wavelength picture of the universe.
The model take into account all important components that concur to the SED
of galaxies at wavelengths from X-rays to the radio. We review the
modeling of each component and provide several applications,
interpreting observations of galaxy of different types at all the wavelengths.
\end{abstract}

\maketitle


\section{Introduction}

The direct study of the process of formation and evolution of galaxies is 
now a realistic aim of observational cosmology.  Thanks to facilities such 
as the HST, ISO, Chandra, XMM-Newton, SCUBA etc, now the theories of galaxy 
formation can be directly tested against multiwavelength observations of
galaxies at different cosmological epochs.

A panchromatic approach is critical both to obtain diagnostics of
the various physical phenomena as well as to estimate the amount
of reprocessing, due to the ISM, of the radiation produced by
primary processes.
In particular, a proper treatment of dust attenuation is
compulsory when estimating the SFR from rest-frame UV luminosities
(e.g. the cosmic SFR from Lyman Break Galaxies). This is
particularly true if we consider that star forming regions are
particularly dense and dusty environments. In a galaxy, dust
reprocessing affects more the youngest stellar population, and the
ensuing picture is much more complex than simple attenuation
models frequently used.
On the other hand, thermal emission from dust in these regions
provides an alternative way to determine the star formation activity.

To interpret the presently available observations,  galaxy
evolution models should include a careful treatment of the above
mentioned effects. In this contribution, we review our population
synthesis model GRASIL\footnote{The code is available at 
\url{http://adlibitum.oat.ts.astro.it/silva/default.html} or
\url{http://web.pd.astro.it/granato/grasil/grasil.html}}, which predicts time dependent
SEDs of galaxies from X-rays to radio, including state-of-the-art
treatment of dust and PAH reprocessing \cite{silv98,gran00}, production of 
radio photons by thermal and non-thermal
processes \cite{bres02}, nebular lines emission \cite{panu03}, 
molecular line emission, and X-rays
expected from stellar populations.


\section{Dust reprocessing}

One of the most important distinctive features of GRASIL is that
it has included, for the first time, the effect of {\it age
selective extinction} of stellar populations (younger stellar
generations are more affected by dust obscuration), mainly (but
not only, see below) due to the fact that stars form in a denser
than average environment, the molecular clouds (MCs), and
progressively get rid of them.

Once the star formation and chemical enrichment histories of a
galaxy are given, GRASIL computes the interaction between the
stellar radiation and dust using a relatively realistic and
flexible geometry for both stars and dust. In general, the system
is described as a superposition of an exponential disk component
and a bulge component, the latter modeled by a King profile. The
ISM is divided into two phases, the diffuse ISM, corresponding to
cirrus dust, and the much denser molecular clouds (MCs). Also, the
fact that new stars are born inside MCs and progressively get rid
of them (either because they escape or because the clouds are
destroyed) is taken into account. This process leads to an {\it
age selective extinction} of stellar generations, in the sense that
the younger the stars, the more they are affected by dust
obscuration in MCs. This is described in GRASIL assuming
that the fraction of starlight radiated inside the clouds by stars
is a function of the star age.
In practice, if $t_{\rm esc}$ is the timescale for the process, 
100\% of the stars younger than $t_{esc}$ are
considered to radiate inside the MCs, and this percentage goes
linearly to 0\% in $2 t_{\rm esc}$. The timescale $t_{\rm esc}$ is a
fundamental parameter, which seems to be quite longer in starbursts
than in normal disk-like galaxies \cite{silv98}.

The starting input for GRASIL is the history
of star formation and chemical enrichment of the system. This is
computed by some external code, which can result from a complex
scenario for the formation of galaxies in a cosmological context
(e.g. \cite{gran01,gran04,baug04}), or a standard chemical evolution
model for the formation of a single galaxy. Here, as in Silva et
al. \cite{silv98}, we follow the latter approach.

The total gas mass (diffuse+MCs) of the galaxy at time $T_{\rm G}$ is
given by the chemical evolution model. The relatively fraction of
molecular gas is a free parameter of the code, $f_{\rm mc}$. The total
molecular mass, $M_{\rm mc}$ is then subdivided into spherical clouds
of mass and radius, $m_{\rm mc}$ and $r_{\rm mc}$. Then, the radiative
transfer of starlight through the MCs and diffuse ISM. The
original version of GRASIL presented in Silva et al. \cite{silv98} used
only a single population of MCs all with the same mass and radius,
the code has been later updated to treat many populations, each
one characterized by its mass $m_{{\rm mc},i}$, radius $r_{{\rm mc},i}$ and
total mass in the population $M_{{\rm mc},i}$ 
(Silva, 1999, PhD thesis \cite{silv99}). 

The ratio between gas mass and dust mass $\delta$ is usually assumed
to scale linearly with the metallicity of the residual gas. This
quantity times $m_{\rm mc}/r_{\rm mc}^2$ determine the optical depth of the
clouds. The predicted SED depend on $m_{\rm mc}$ and $r_{\rm mc}$ only through
the combination $m_{\rm mc}/r_{\rm mc}^2$, which is the true free parameter.

For the dust composition a mixture of graphite and silicate
grains and PAH is adopted, with size distributions triggered to
match the extinction and emissivity properties of the local ISM
(for more details see Silva et al. \cite{silv98}). However the code has the
flexibility to modify the dust mixture, and to differentiate
between MCs and cirrus.

The SSPs included in GRASIL are based on the Padova stellar
models, and cover a large range of ages and metallicities.
Starlight reprocessing from dust in the envelopes of AGB stars is
included directly into the SSPs, as described by Bressan et al. \cite{bres98}.


\section{Radio emission}

The existence of a correlation between FIR and Radio emission
is locally well established over a significant range of luminosity, from
normal spirals to the most extreme ULIRGs. At 1.49 GHz, 
$q_{1.49}=\log (F_{\rm FIR}/F_{1.49\rm GHz}/\nu_{\rm FIR}) \simeq 2.35\pm 0.2$  
\cite{sand96}.
This correlation suggests that non-thermal (NT) emission is related to
the recent star formation and the most likely mechanism is
synchrotron emission from relativistic electrons accelerated into
the shocked interstellar medium, following core collapsed
supernova (CCSN) explosions \cite{cond90}. Its small
scatter testifies the universal proportions with which energy is
radiated away at IR and radio wavelengths. Its validity has been
recently confirmed up to redshift $\simeq 1.3$ \cite{garr02} and
it is widely extrapolated much beyond, to estimate the redshift of
more distant objects (e.g. \cite{cari00}).

\begin{figure}
\includegraphics[width=0.5\textwidth]{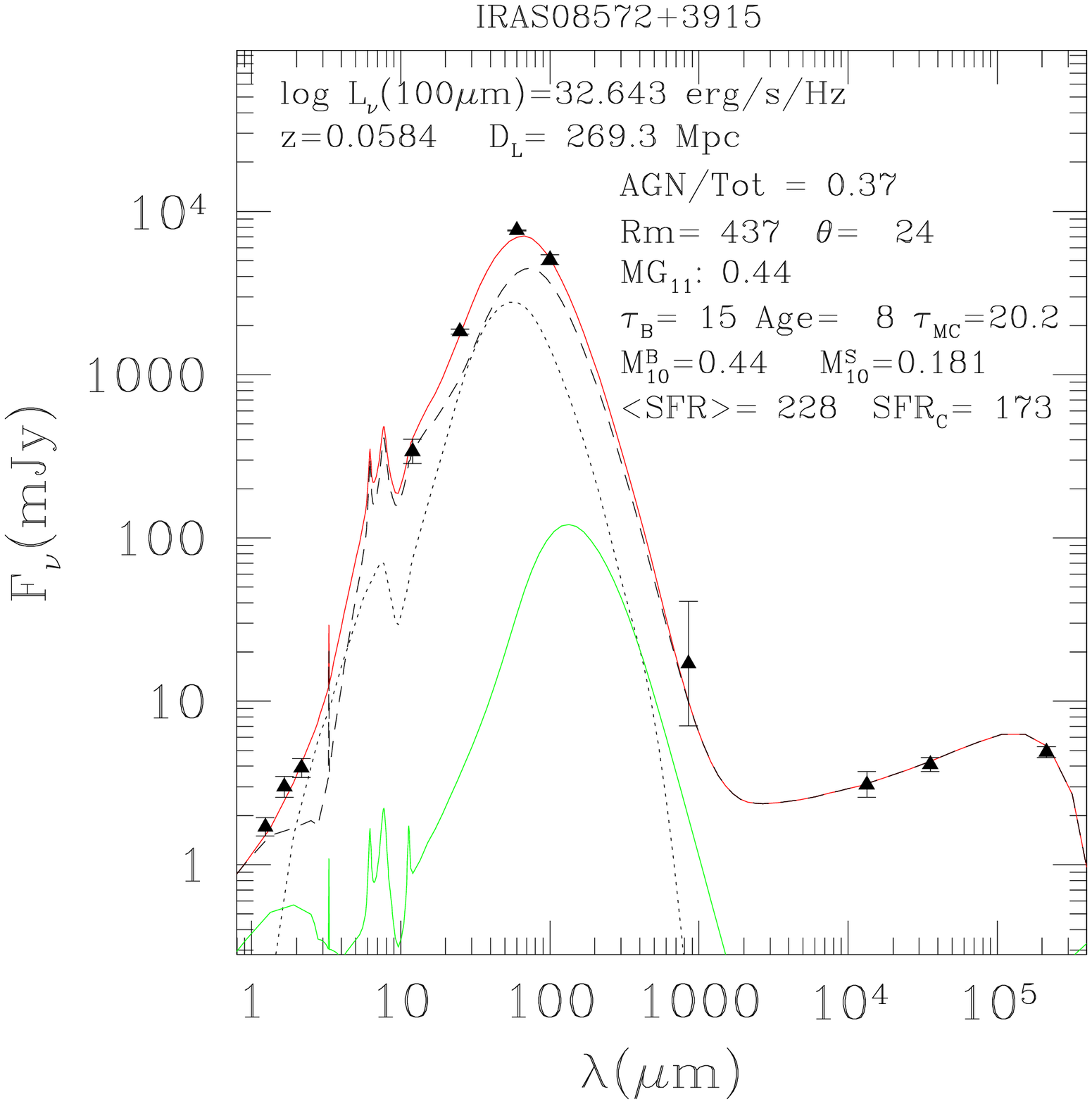}
\includegraphics[width=0.5\columnwidth]{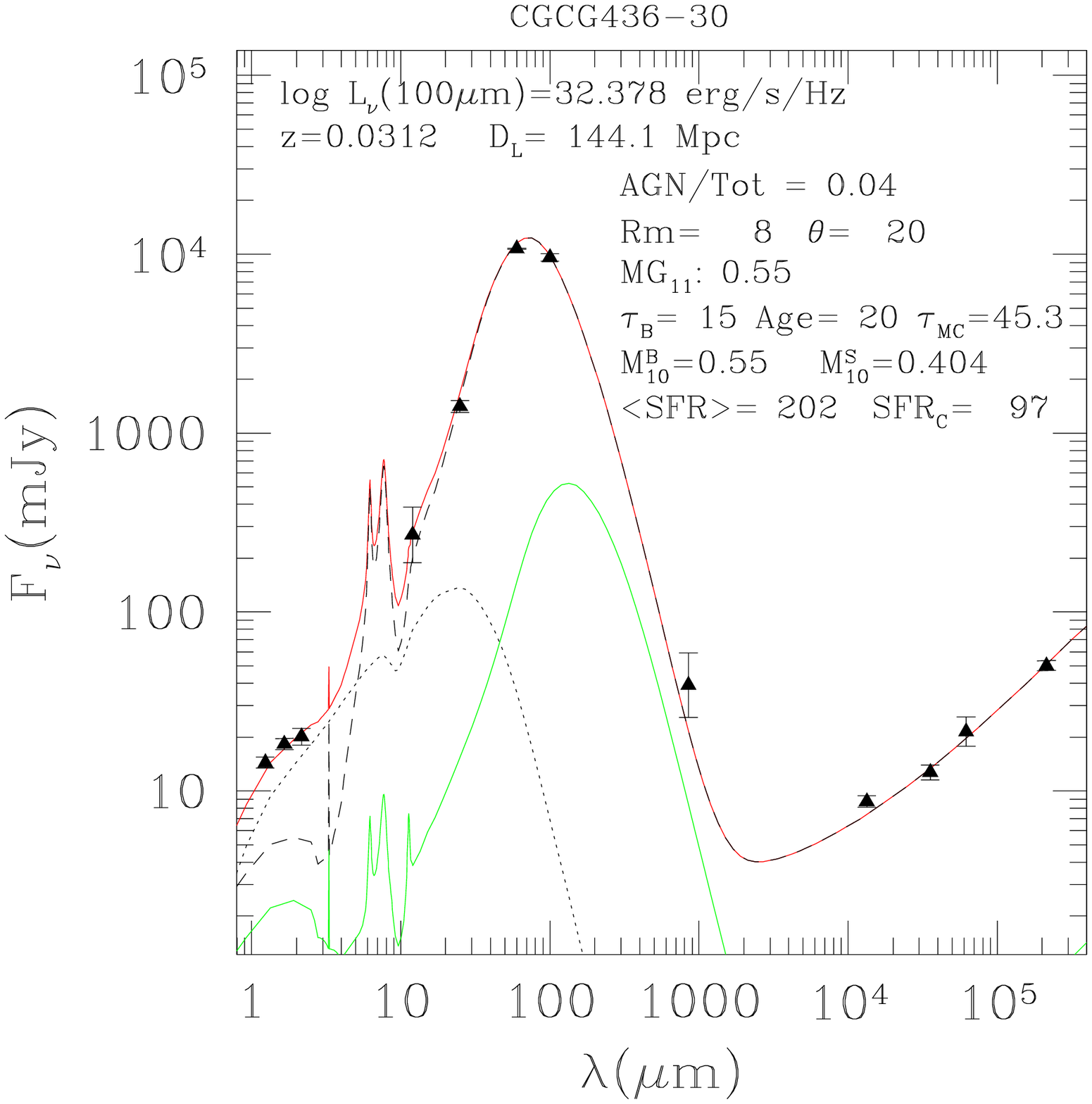}
\caption{NIR to radio SED fits of two compact ULIRGs,
IRAS08572+3915 and CGCG436-30. IRAS08572+3915 has a "flat" radio
slope, due to  free-free emission from HII regions. This is
typical of "young" starburst. Its 1.4GHz flux is attenuated by
free-free absorption. The radio SED of CGCG436-30 is already
dominated by synchrotron emission from CCSNs, a feature typical of
older starbursts. Upper solid line is the SED fit, dashed line is
the contribution from the galaxy (starburst plus disk), dotted
line is the contribution from the AGN and the lower solid line is
the contribution of the disk alone. See Prouton et al. \cite{prou04} for
details.} \label{inter}
\end{figure}

In order to increase the diagnostic capabilities of GRASIL,
Bressan, Silva \& Granato \cite{bres02} have extended its domain
into the radio wavelengths. In brief, two main components were
considered, the thermal emission from \hii\ regions and synchrotron
emission from relativistic electrons. The thermal component
($L_\nu^{\rm T}$), is proportional to the number of hydrogen
ionizing photons ($Q_{\rm H}$) derived from the adopted SSPs:
\begin{equation}
L_\nu^{\rm T}\propto Q_{\rm H} T_e^{0.45} \nu^{-0.1}
\end{equation}
The non-thermal ($L_\nu^{\rm NT}$) component has been calibrated 
with the NT emission and the CCSN rate ($r_{\rm CCSN}$)
of our Galaxy, after accounting for the small contribution, 
$\simeq$6\%, of radio supernova remnants ($E^{\rm SNR}$):
\begin{equation}
L_\nu^{\rm NT}\propto (E^{\rm SNR}\nu^{-0.5} + E^{el}\nu^{-0.9})r_{\rm CCSN}
\end{equation}
Though the calibration with our galaxy is quite uncertain, with
these assumptions Bressan et al. \cite{bres02} and Panuzzo et al. \cite{panu03}
were able to reproduce well the FIR/Radio correlation of normal
star forming galaxies, namely $q_{1.49}\simeq$2.35. Bressan et
al. \cite{bres02} have also shown that, due to different spectral index of
thermal and NT radio emission and the delay of CCSN explosions,
radio emission can be used to analyze obscured starbursts with a
time resolution of a few tens of Myr, unreachable with other star
formation indicators. This is shown in Figure \ref{inter},
from Prouton et al. \cite{prou04}. A systematic study of
compact ULIRGs from NIR to radio wavelengths is underway (Clemens
et al. in preparation). This will provide important clues on the
contribution of the AGN in ULIRGs and on the technique of
photometric redshift determination of high-z obscured sources.


\section{Molecular lines}

The emission from molecular lines is used to diagnose the physical
conditions within star forming molecular clouds. In fact, emission
lines such as those from CS, HCN and HCO$^+$ are excited only in
high density gas, while others 
(e.g. $^{12}$CO) trace more typical regions in the cold ISM. The study
of the physical conditions of the star formation regions, the star
formation efficiencies, the quantification of the contamination of
molecular lines to the sub-mm continuum, and the possibility of
obtaining spectroscopic redshifts of distant galaxies using
molecular lines are some of the challenges of the next future
millimeter astronomy. Therefore, the joint modeling of molecular
and dust emissions is essential to investigate the global
relationships between gas, dust and star formation.

We have developed a method that combines GRASIL and a molecular lines
emission model, based on LVG approximation (Vega et al.\ in
preparation).
Our LVG code (see de Jong et al. \cite{dejo75} for the mathematical
formulation) takes into account collisional excitation by H$_2$
molecules, radiative trapping and radiative excitation  by the
cosmic microwave background. The adjustable LVG parameters are:
the numerical density of colliders, $n_{\rm H_{2}}$, the gas kinetic
temperature, $T_{\rm K}$, and the parameter 
$\Lambda= X_{\rm mol} r_{\rm mc}/\Delta V$ in  pc km$^{-1}$, 
where $X_{\rm mol}$ is the relative abundance of
a given molecule to H$_2$. Our LVG code at present treats
$^{12}$CO, $^{13}$CO, C$^{18}$O, CS, HCN, HNC and HCO$^+$.

The main advantage of combining the LVG code with GRASIL for the
analysis of molecular emission in galaxies is that, by fitting the
galaxy SED from the far UV to the radio bands, GRASIL can provide
realistic estimates of the SFR, fraction of molecular gas, gas to
dust ratio, dust temperature, and the {\it average} molecular cloud
structure.

The shape of the predicted IR SED depends mainly on the
optical depth of the clouds, $\tau_{\rm mc} \propto \delta\cdot m_{\rm mc}/r_{\rm mc}^2$. 
After $\tau_{\rm mc}$ is set from the fit to the continuum, we obtain
the molecular gas density
$n_{\rm H_{2}}=0.7 m_{\rm mc}/(4 \pi r_{\rm mc}^3 m_{\rm H_{2}}/3)$
corresponding to different $m_{\rm mc}$ (within reasonable values),
assuming that most of the molecular emission comes from the same clouds with
ongoing star formation. We apply our LVG model to the clouds for each
value of $m_{\rm mc}$ (or $n_{\rm H_{2}}$), adjusting $T_{\rm K}$ and 
$\Lambda$ to fit the observed intensity of the molecular lines lines. 
The best fit model gives the best estimate of the excitation conditions of
molecular gas.

\begin{figure}
\includegraphics[width=0.55\textwidth]{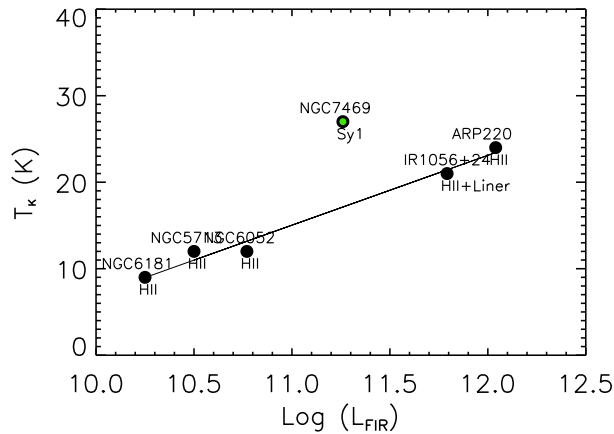}
\caption{Kinetic temperature versus FIR luminosity.
The line represents the
linear fit for the starburst galaxies.} \label{fig:olga}
\end{figure}

As an example, we show some results of applying our GRASIL+LVG
model to six galaxies, four LIRGs and two ULIRGs, for which both
continuum and molecular line data are available. From our
preliminary analysis, there are hints that $T_{\rm K}$  is tightly
correlated to the FIR (and Radio) emission over a wide range of
luminosity (Figure \ref{fig:olga}). The only galaxy out of this
correlations is a Sy1.2. A likely explanation is that, in
starburst galaxies, the molecular excitation is driven by
processes related to star formation. In the case of the AGN, the
molecular gas seems to be ``overheated'', probably by the central
engine. The mean molecular densities found for these galaxies are $<10^{3.5}$
cm$^{-3}$ and the bulk of CO emission is coming from a low density medium. 
However, observations of HCN suggest the presence of a much denser
medium, thus we are working in a two-zone model:
dense clouds embedded in a low density medium.


\section{X-ray emission from galaxies}

The X-ray range has recently become very effective to study galaxy
evolution, thanks to the high resolution, low flux levels and
large spectral range reached by {\it Chandra} and {\it
XMM-Newton}, that allow to probe the faint X-ray sources of star
forming galaxies.
In several works a strong correlation between the X-ray and the
FIR or radio luminosity has been reported for star forming
galaxies (e.g. \cite{grif90,davi92,helf01,rana03}).
Grimm et al. \cite{grim03}, by comparing the luminosity functions of
X-ray point sources (XLF) of several galaxies, conclude that 
the shape of the XLF being almost universal.
Therefore, the X-ray emission can be considered another
independent indicator of star formation rate and history, to
exploit together with the radio, IR and UV indicators, in order to
set new constraints for a multi-wavelength tackling of galaxy
evolution studies.
To complete our self consistent modeling of the SED of galaxies,
we have extended GRASIL to the X-ray range (\cite{silv03} and Silva et al.
in preparation).

We consider the following sources (see \cite{vanb00,pers02}): 
(1) Binaries in which the donor is
a core H-burning OB star that fills its Roche lobe, and the
primary is a black hole (BH) or a Neutron Star (NS) (hereafter RBH
and RNS); (2) Binaries in which the donor is a shell H-burning OB
star undergoing a phase of strong stellar wind, and the primary is
a BH or a NS (hereafter WBH and WNS); (3) Pulsars (PLS); (4)
Supernova remnants (SNRs); (5) low-mass X-rays binaries (LMXRBs).

The X-ray luminosity from the different components is directly
included in the library of Simple Stellar Population.
We proceed in the following way:
(i) at each age of a stellar population we know the number of
stars in the different evolutionary phases or dead remnants,
therefore we compute how many stars of main sequence or post main
sequence combine to form binary systems with BH or NS as a
primary; (ii) then each source is assigned the appropriate $L_X$,
therefore we obtain the total $L_X$ of a population of objects of
each kind; (iii) we distribute this energy according to suitable
(observed average) spectral shapes.

The inputs needed for the computation are the following (see 
\cite{dedo98,vanb00}): the IMF for single and primary stars, the 
distribution of the initial mass ratio $M_2/M_1$, the binary 
frequency at birth ($\geq 0.5$), the distribution over the periods.
Of course the results depend also on assumptions on stellar evolution 
(mainly mass loss rate, remnant masses, initial masses for BH 
formation, dependence on metallicity).
As for LMXRB, they are linked to the old stellar populations, i.e.
to the mass content of galaxies \cite{grim03}. We adopt the
ratio $L_X/M_*$ evaluated for LMXRBs in our Galaxy by Grimm et
al. ($5\times 10^{28}$ erg/s/M$_\odot$) and assume this ratio is
the same in all galaxies.

\begin{figure}
\includegraphics[width=0.5\textwidth]{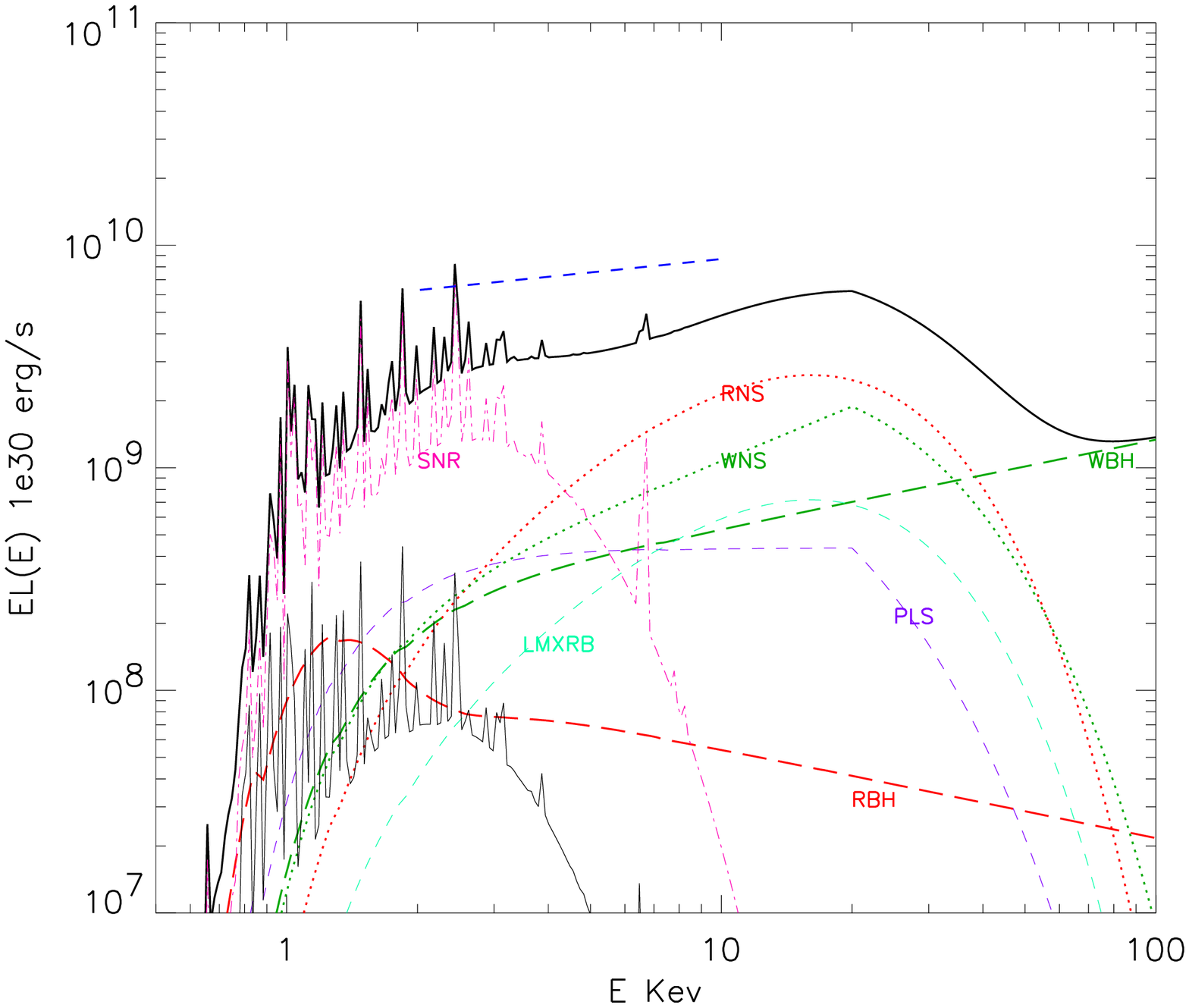}
\includegraphics[width=0.5\textwidth]{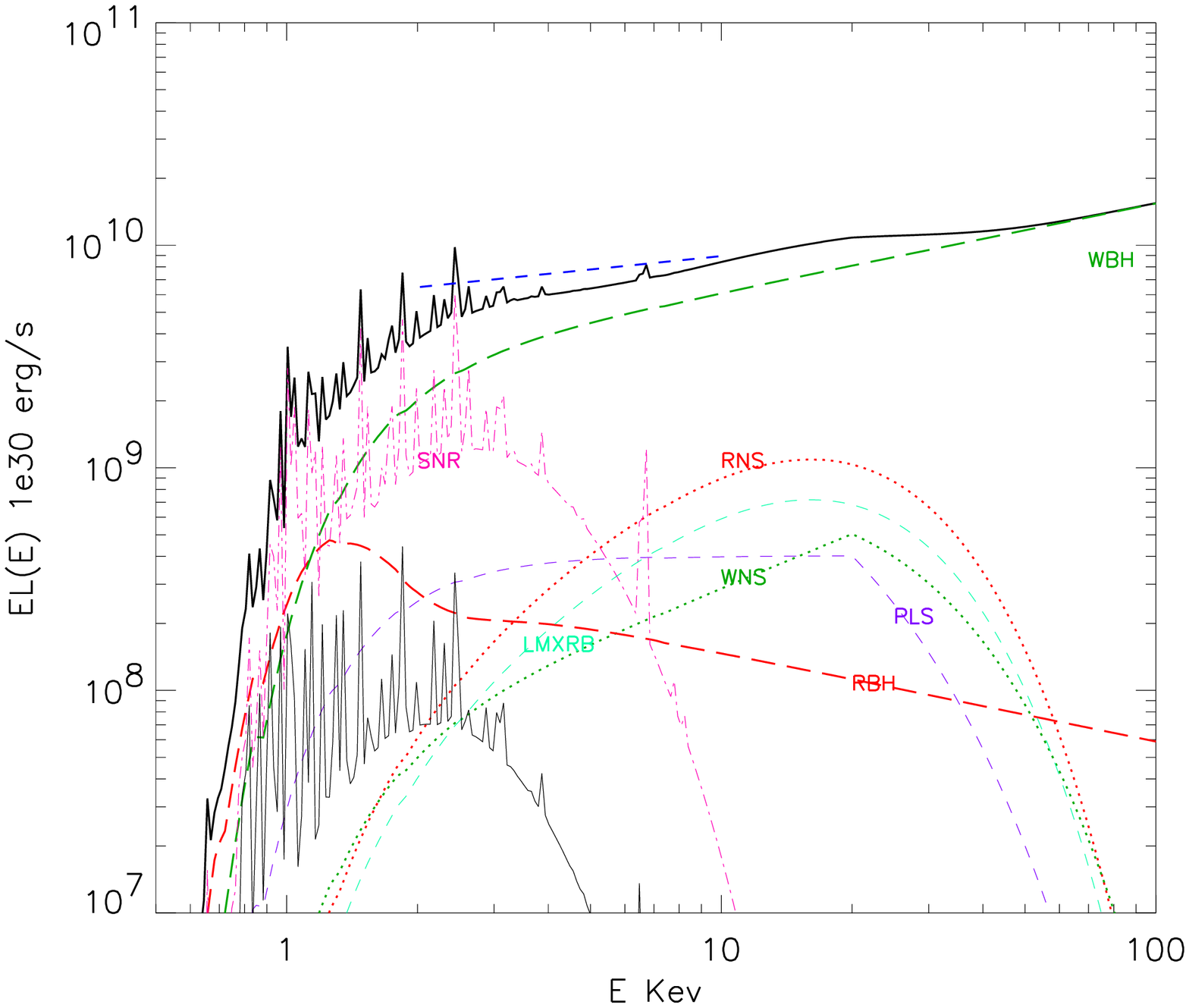}
\caption{X-ray SED for M82, obtained from the SF history of the
model fitting the UV to radio data. In the first plot we have used
the X-ray predicted by including the SSPs of all the metallicities
(from $Z=0.0004$ to $0.05$), while in second plot we have used
only the $Z=0.008$ SSP. In the latter case, even though the wind
mass loss rate is lower, the higher BH remnant masses determine
much higher accretion luminosities. Note however that the stellar
inputs (initial-final mass relation, its dependence on
metallicity, initial mass for the formation of a BH...) are very
uncertain. Here an absorbing column density $N_H=10^{22}$
cm$^{-2}$ has been applied to the model SED. The observed SED
(dashed line from 2 to 10 kev) is from Griffiths et al. \cite{grif00}.}
\label{xm82}
\end{figure}

Once we have the SED of the populations of different X-ray sources
as a function of age and metallicity, the integrated X-ray SED for
a model galaxy of age $T_{\rm G}$ is given by:
\begin{equation}
L_X(T_{\rm G},\epsilon)= \sum_j \int_0^{T_{\rm G}} \! \! f_j(\epsilon)
L_{X,j}^{SSP}(T_{\rm G}-t,Z(t))\, \psi(t)\, dt \label{eq:syn}
\end{equation}
where $\psi(t)$ and $Z(t)$ are the star formation rate and metallicity
enrichment history for the model, $j$ refers to the different X-ray
sources, and $f_j(\epsilon)$ is the spectral shape associated to
source of kind $j$.
In Fig. \ref{xm82} we show the X-ray SED resulting from our model
for M82. The star formation and metallicity evolution for M82 are obtained by
fitting its UV to radio SED \cite{silv98,bres02}. 
The plots show the important effect on assumptions on
stellar evolution and metallicity. Work is in progress to explore
these effects, also connected with the presence of the so called
Ultra-Luminous X-ray sources (ULX) in star forming galaxies (e.g.
\cite{gilf04}). These sources are interpreted as intermediate mass 
BHs ($M >100$ M$_\odot$) or to specific effects of the accretion onto 
normal BHs \cite{bege02,rapp04} with a violation of the strict Eddington limit.


\section{nebular emission lines}

Emission lines from \hii\ regions are a powerful diagnostic of the
instantaneous SFR because the luminosity of H recombination lines is
proportional to the number of massive, short-living stars. Moreover,
emission lines are widely used to estimate the dust attenuation, e.g.
with the Balmer Decrement.
Nebular lines are also used to estimate the metallicity of
the gas and, finally, to infer the nature and age of the ionizing source.
For all the above reasons, we implemented the computation of nebular 
lines in GRASIL (see Panuzzo et al. \cite{panu03} for details).

To compute the line emission intensities in a population synthesis model, one should consider
the spectrum of the ionizing source, provided by the recent star formation history, and use a
photoionization code with suitable values of the gas parameters.
It becomes particularly time consuming in applications requiring a
large number of models. Our approach has been to pick out the physical
parameters which actually affect the emission properties of \hii\ regions.

The emission spectrum from a single \hii\ region depends on two
main ingredients: the SED of the ionizing star cluster (determined by the 
IMF and the total mass, age and metallicity of the cluster, as well as 
by the adopted model atmospheres) and the properties of the excited gas 
(the density, the chemical composition and the geometry). 

We found that the emission line spectrum of an \hii\ region with fixed gas
properties is described with reasonable precision by only three quantities: 
the number of ionizing photons per second for \ion{H}{i}, \ion{He}{i} and \ion{O}{ii}
($Q_{\mathrm H}$, $Q_{\mhe}$, and $Q_{\mathrm O}$) emitted by the
ionizing source.
We used as ionizing SEDs analytical spectra composed by piece-wise blackbodies,
similar to young SSP spectra, with temperatures given as functions
of $Q_{\mathrm H}$, $Q_{\mhe}$, and $Q_{\mathrm O}$.
We shown that our analytical spectra produce the same emission line spectra 
of SSP with the same values of $Q_{\mathrm H}$, $Q_{\mhe}$, and 
$Q_{\mathrm O}$, within a reasonable accuracy. Thus, we computed a library 
of photoionization models as a function of $Q_{\mathrm H}$, $Q_{\mhe}$, 
and $Q_{\mathrm O}$. When estimating the actual line emission due to a given 
stellar population, the spectrophotometric code computes
$Q_{\mathrm H}$, $Q_{\mhe}$, and $Q_{\mathrm O}$ from the corresponding
SED, and then interpolates the value from the above library.

The library was computed for a grid of values of
$Q_{\rm H}$, $Q_{\rm He}/Q_{\rm H}$, and $Q_{\rm O}/Q_{\rm He}$,
for different assumptions on the gas
density, metallicity and filling factor $\epsilon$.
This method allows to get rid of the particular SSP model and IMF. Moreover
it can be implemented in every population synthesis code. The library of
photoionization models is available to the astronomical community from a
web site\footnote{\url{http://web.pd.astro.it/panuzzo/hii}}.

\begin{figure}
\includegraphics[angle=270,width=0.5\textwidth]{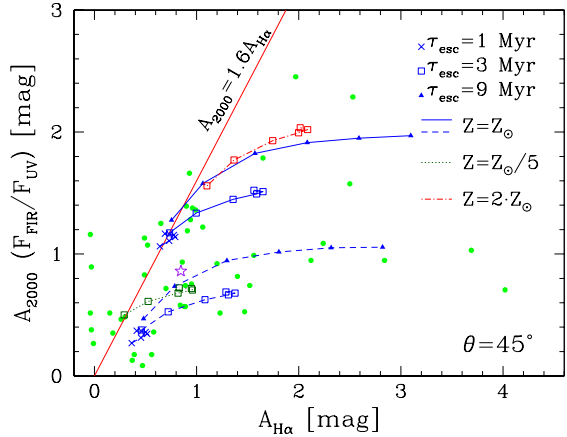}
\caption{The $A_{\rm UV}$ derived from the ratio $F_{\rm FIR}/F_{\rm UV}$
vs. $A_{\rm H\alpha}$ from Balmer decrement.
Filled circles show Buat et al. \cite{buat02} sample, compared with models 
(different symbols connect by lines). The $A_{\rm UV}=1.6A_{\rm H\alpha}$ is 
the observed correlation for starbursts. See Panuzzo et al. \cite{panu03} for details.}
\label{auvaha}
\end{figure}

A simple application of our population synthesis with nebular emission
and dust processing was to study different methods
to estimate the attenuation in normal spirals \cite{panu03}.
UV attenuation ($A_{\rm UV}$) can be estimated by $F_{\rm FIR}/F_{\rm UV}$ ratio
(see \cite{meur99,hira03}) while the attenuation suffered by H$\alpha$ ($A_{\rm H\alpha}$) 
can be derived from the Balmer decrement.
Observing a sample of normal spiral galaxies, Buat et al. \cite{buat02} found that
$A_{\rm UV}$ and $A_{\rm H\alpha}$ do not correlate, contrary to what found in starbursts 
galaxies \cite{calz97,buat02}.

Our models \cite{panu03} (see Fig. \ref{auvaha}) reproduce location of the
observed galaxies and confirm a real lack of correlation between the $A_{\rm UV}$ 
from $F_{\rm FIR}/F_{\rm UV}$ and $A_{\rm H\alpha}$.
The dispersion is a clear effect of the age selective extinction. In fact 
massive ionizing stars have a short lifetime so they can spend all their life
inside dusty environments (i.e. MCs) while UV emitting stars that have a longer 
lifetime can radiate from a less dusty medium.


\section{PAH modeling}

The mid infrared (MIR) spectra of many dusty galactic and
extragalactic objects show strong emission features
most commonly ascribed to aromatic C-C and C-H vibrations in large
planar Polycyclic Aromatic Hydrocarbons (PAH) molecules (e.g. \cite{lege84}).

Through a comprehensive analysis of the MIR to FIR spectra of
Galactic star forming regions, normal and starburst galaxies and
AGN, Peeters et al. \cite{peet04} have shown that the SED of galaxies is
very similar to that of Galactic exposed PDRs, and that PAHs are a
better tracer of B stars rather than by massive star formation (O
stars). This has been found also by Tacconi-Garman et al. \cite{tacc04}
by direct high resolution $3 \mu$m imaging of two nearby starbursts
(NGC253 and NGC1808): they find no spacial coincidence between the
detailed distribution of PAH emission and the locations of most
recent star formation.

On the theoretical side, the new observations have allowed to
better define the modeling of PAHs (e.g. \cite{lidr01,zubk04}). 
In order to interpret at best the spectra of
galaxies, a detailed modeling that includes the complex
interaction between stars and dust is needed. To this aim we have
updated GRASIL by including a better treatment of PAH emission
bands. The previous version of GRASIL \cite{silv98,silv99} included PAH
modeling based on pre-ISO data: the five $3$ to $11 \mu$m PAH
bands were computed following mainly Xu \& De Zotti \cite{xude89}. 
We have now updated our treatment according to the PAH
model by Li \& Draine \cite{lidr01} (LD01). In particular: (i) ISO
observations have increased the number of known bands, therefore
now in addition to the $3.3$, $6.2$, $7.7$, $8.6$, $11.3 \mu$m
bands, we compute the $11.9$, $12.7$, $16.4$, $18.3$, $21.2$, and
$23.1 \mu$m bands; (ii) we adopt a Drude profiles for the
features; (iii) the FWHM and emission cross section for the new
bands are from LD01.
We have compared our updated PAH model with ISO spectra of
galaxies by Lu et al. \cite{luhe03} in order to test weather the LD01
model, suited for the diffuse ISM, can be safely adopted in
general for star forming galaxies. We searched continuum NIR to
radio data for the galaxies of the Lu et al. sample, and fitted the
continuum SED. We then made a detailed comparison of the
corresponding model PAH with the Lu et al. MIR spectra. In all
cases the predicted PAHs compare very well with the observation,
the greatest disagreements are always within a factor of 1.5, but
typically much less.


\section{conclusions}

GRASIL is a realistic model for simulating detailed and panchromatic
SEDs of galaxies. We are constantly working to improve GRASIL modeling in particular
about PAH, X-rays, molecular line emission, and dust geometry.
The model was successfully used in many works and in several different fields, 
also by researchers outside the developing team.
GRASIL is available for the scientific community: everybody
is welcome to download the code from our web site or to ask for the source.
We are now working on a web interface: any researcher will be able to run
GRASIL on our computers, and to get the result by e-mail.


\begin{theacknowledgments}
We thank many others with whom we have had constructing discussions. 
We also thank A. Petrella and L. Paoletti for their effort in developing 
the GRASIL web interface. P.P. acknowledges support from ASI project
``Cosmologia e Fisica fondamentale dallo Spazio''. 
\end{theacknowledgments}


\bibliographystyle{aipproc}
\bibliography{biblio}

\begin{thebibliography}{39}
\expandafter\ifx\csname natexlab\endcsname\relax\def\natexlab#1{#1}\fi
\providecommand{\enquote}[1]{``#1''}
\expandafter\ifx\csname url\endcsname\relax
  \def\url#1{\texttt{#1}}\fi
\expandafter\ifx\csname urlprefix\endcsname\relax\def\urlprefix{URL }\fi

\bibitem[Silva et~al.(1998)]{silv98}
Silva, L., Granato, G.~L., Bressan, A., and Danese, L., \emph{ApJ},
  \textbf{509}, 103 (1998).

\bibitem[Granato et~al.(2000)]{gran00}
Granato, G.~L., Lacey, C.~G., Silva, L., Bressan, A., Baugh, C.~M., Cole, S.,
  and Frenk, C.~S., \emph{ApJ}, \textbf{542}, 710 (2000).

\bibitem[Bressan et~al.(2002)]{bres02}
Bressan, A., Silva, L., and Granato, G.~L., \emph{A\&A}, \textbf{392}, 377
  (2002).

\bibitem[Panuzzo et~al.(2003)]{panu03}
Panuzzo, P., Bressan, A., Granato, G.~L., Silva, L., and Danese, L.,
  \emph{A\&A}, \textbf{409}, 99 (2003).

\bibitem[Granato et~al.(2001)]{gran01}
Granato, G.~L., Silva, L., Monaco, P., Panuzzo, P., Salucci, P., Zotti, G.~D.,
  and Danese, L., \emph{MNRAS}, \textbf{324}, 757 (2001).

\bibitem[Granato et~al.(2004)]{gran04}
Granato, G.~L., {De Zotti}, G., Silva, L., Bressan, A., and Danese, L.,
  \emph{ApJ}, \textbf{600}, 580 (2004).

\bibitem[Baugh et~al.(2004)]{baug04}
Baugh, C.~M., Lacey, C.~G., Frenk, C.~S., Granato, G.~L., Silva, L., Bressan,
  A., Benson, A.~J., and Cole, S., \emph{MNRAS} (2004), in press
  (astro-ph/0406069).

\bibitem[Silva(1999)]{silv99}
Silva, L., \emph{Modelling the SED Evolution of Dusty Galaxies and
  Applications}, Ph.D. thesis, SISSA (1999).

\bibitem[Bressan et~al.(1998)]{bres98}
Bressan, A., Granato, G.~L., and Silva, L., \emph{A\&A}, \textbf{332}, 135
  (1998).

\bibitem[Sanders and Mirabel(1996)]{sand96}
Sanders, D.~B., and Mirabel, J.~F., \emph{ARA\&A}, \textbf{34}, 749 (1996).

\bibitem[Condon and Yin(1990)]{cond90}
Condon, J.~J., and Yin, Q.~F., \emph{ApJ}, \textbf{357}, 97 (1990).

\bibitem[Garrett(2002)]{garr02}
Garrett, M.~A., \emph{A\&A}, \textbf{384}, L19 (2002).

\bibitem[Carilli and Yun(2000)]{cari00}
Carilli, C.~L., and Yun, M.~S., \emph{ApJ}, \textbf{530}, 618 (2000).

\bibitem[Prouton et~al.(2004)]{prou04}
Prouton, O.~R., Bressan, A., Clemens, M., Franceschini, A., Granato, G.~L., and
  Silva, L., \emph{A\&A}, \textbf{421}, 115 (2004).

\bibitem[{de Jong} et~al.(1975)]{dejo75}
{de Jong}, T., Dalgarno, A., and Chu, S., \emph{ApJ}, \textbf{199}, 69 (1975).

\bibitem[Griffiths and Padovani(1990)]{grif90}
Griffiths, R.~E., and Padovani, P., \emph{ApJ}, \textbf{360}, 483 (1990).

\bibitem[David et~al.(1992)]{davi92}
David, L.~P., Jones, C., and Forman, W., \emph{ApJ}, \textbf{388}, 82 (1992).

\bibitem[Helfand and Moran(2001)]{helf01}
Helfand, D.~J., and Moran, E.~C., \emph{ApJ}, \textbf{554}, 27 (2001).

\bibitem[Ranalli et~al.(2003)]{rana03}
Ranalli, P., Comastri, A., and Setti, G., \emph{A\&A}, \textbf{399}, 39 (2003).

\bibitem[Grimm et~al.(2003)]{grim03}
Grimm, H., Gilfanov, M., and Sunyaev, R., \emph{MNRAS}, \textbf{339}, 793
  (2003).

\bibitem[Silva et~al.(2003)]{silv03}
Silva, L., Granato, G.~L., Bressan, A., and Panuzzo, P., \enquote{Modelling the
  Radio to X-ray SED of Galaxies,} in \emph{Galaxy Evolution: Theory \&
  Observations}, edited by V.~Avila-Reese, C.~Firmani, C.~S. Frenk, and
  C.~Allen, Revista Mexicana de Astronom{\'\i}a y Astrof{\'\i}sica, 2003,
  vol.~17, p.~93.

\bibitem[{Van Bever} and Vanbeveren(2000)]{vanb00}
{Van Bever}, J., and Vanbeveren, D., \emph{A\&A}, \textbf{358}, 462 (2000).

\bibitem[Persic and Rephaeli(2002)]{pers02}
Persic, M., and Rephaeli, Y., \emph{A\&A}, \textbf{382}, 843 (2002).

\bibitem[{de Donder} and Vanbeveren(1998)]{dedo98}
{de Donder}, E., and Vanbeveren, D., \emph{A\&A}, \textbf{333}, 557 (1998).

\bibitem[Griffiths et~al.(2000)]{grif00}
Griffiths, R.~E., Ptak, A., Feigelson, E.~D., Garmire, G., Townsley, L.,
  Brandt, W.~N., Sambruna, R., and Bregman, J.~N., \emph{Science},
  \textbf{290}, 1325 (2000).

\bibitem[Gilfanov(2004)]{gilf04}
Gilfanov, M., \emph{MNRAS}, \textbf{349}, 146 (2004).

\bibitem[Begelman(2002)]{bege02}
Begelman, M.~C., \emph{ApJ}, \textbf{568}, L97 (2002).

\bibitem[Rappaport et~al.(2004)]{rapp04}
Rappaport, S.~A., Podsiadlowski, P., and Pfahl, E., \emph{MNRAS} (2004), in
  press (astro-ph/0408032).

\bibitem[Buat et~al.(2002)]{buat02}
Buat, V., Boselli, A., Gavazzi, G., and Bonfanti, C., \emph{A\&A},
  \textbf{383}, 801 (2002).

\bibitem[Meurer et~al.(1999)]{meur99}
Meurer, G.~R., Heckman, T.~M., and Calzetti, D., \emph{ApJ}, \textbf{521}, 64
  (1999).

\bibitem[Hirashita et~al.(2003)]{hira03}
Hirashita, H., Buat, V., and Inoue, A.~K., \emph{A\&A}, \textbf{410}, 83
  (2003).

\bibitem[Calzetti(1997)]{calz97}
Calzetti, D., \emph{AJ}, \textbf{113}, 162 (1997).

\bibitem[Leger and Puget(1984)]{lege84}
Leger, A., and Puget, J.~L., \emph{A\&A}, \textbf{137}, L5 (1984).

\bibitem[Peeters et~al.(2004)]{peet04}
Peeters, E., Spoon, H. W.~W., and Tielens, A. G. G.~M., \emph{ApJ},
  \textbf{613}, 986 (2004).

\bibitem[{Tacconi-Garman} et~al.(2004)]{tacc04}
{Tacconi-Garman}, L.~E., Sturm, E., Lehnert, M., Lutz, D., Davies, R.~I., and
  Moorwood, A. F.~M., \emph{A\&A} (2004), accepted (astro-ph/0411272).

\bibitem[Li and Draine(2001)]{lidr01}
Li, A., and Draine, B.~T., \emph{ApJ}, \textbf{554}, 778 (2001).

\bibitem[Zubko et~al.(2004)]{zubk04}
Zubko, V., Dwek, E., and Arendt, R.~G., \emph{ApJS}, \textbf{152}, 211 (2004).

\bibitem[Xu and {de Zotti}(1989)]{xude89}
Xu, C., and {de Zotti}, G., \emph{A\&A}, \textbf{225}, 12 (1989).

\bibitem[Lu et~al.(2003)]{luhe03}
Lu, N., Helou, G., Werner, M.~W., Dinerstein, H.~L., Dale, D.~A., Silbermann,
  N.~A., Malhotra, S., Beichman, C.~A., and Jarrett, T.~H., \emph{ApJ},
  \textbf{588}, 199 (2003).

\end{thebibliography}

\end{document}